\documentclass[superscriptaddress,aps,prx,twocolumn,showpacs,nofootinbib, longbibliography, notitlepage,floatfix]{revtex4-2}
\usepackage{etex}
\usepackage{amsmath,amssymb,amsthm}
\usepackage[colorlinks=true,citecolor=blue,urlcolor=blue]{hyperref}
\usepackage[pdftex]{graphicx}
\usepackage{times,txfonts}
\usepackage{braket}
\usepackage{color}
\usepackage{natbib}
\usepackage{amsmath,blkarray}
\usepackage{mathtools}
\usepackage{ulem}
\usepackage{latexsym}
\usepackage{tabularx, booktabs}
\usepackage{graphics,epstopdf}
\usepackage{graphicx}
\usepackage{float}
\usepackage{amsfonts}
\usepackage{tikz}
\usetikzlibrary{quantikz}
\usepackage{color,soul}

\newcommand{\be}{\begin{equation}}
\newcommand{\ee}{\end{equation}}
\newcommand{\ba}{\begin{eqnarray}}
\newcommand{\ea}{\end{eqnarray}}

\usepackage{multirow}
\usepackage{appendix}
\usepackage{url}

\begin{document}
\title{Digital-analog quantum convolutional neural networks for image classification} 

\author{Anton Simen}
\affiliation{Kipu Quantum, Greifswalderstrasse 226, 10405 Berlin, Germany}

\author{Carlos Flores-Garrigos}
\affiliation{IDAL, Electronic Engineering Department, ETSE-UV, University of Valencia, Avgda. Universitat s/n, 46100 Burjassot, Valencia, Spain}

\author{Narendra N. Hegade}
\email{narendrahegade5@gmail.com}
\affiliation{Kipu Quantum, Greifswalderstrasse 226, 10405 Berlin, Germany}

\author{Iraitz Montalban}
\affiliation{Kipu Quantum, Greifswalderstrasse 226, 10405 Berlin, Germany}

\author{Yolanda Vives-Gilabert}
\email{yolanda.vives@uv.es}
\affiliation{IDAL, Electronic Engineering Department, ETSE-UV, University of Valencia, Avgda. Universitat s/n, 46100 Burjassot, Valencia, Spain}

\author{Eric Michon}
\affiliation{Kipu Quantum, Greifswalderstrasse 226, 10405 Berlin, Germany}

\author{Qi Zhang}
\affiliation{Kipu Quantum, Greifswalderstrasse 226, 10405 Berlin, Germany}

\author{Enrique Solano}
\email{enr.solano@gmail.com }
\affiliation{Kipu Quantum, Greifswalderstrasse 226, 10405 Berlin, Germany}

\author{José D. Martín-Guerrero}
\email{jose.d.martin@uv.es}
\affiliation{IDAL, Electronic Engineering Department, ETSE-UV, University of Valencia, Avgda. Universitat s/n, 46100 Burjassot, Valencia, Spain}
\affiliation{Valencian Graduate School and Research Network of Artificial Intelligence (ValgrAI)}

\begin{abstract}
We propose digital-analog quantum kernels for enhancing the detection of complex features in the classification of images. We consider multipartite-entangled analog blocks, stemming from native Ising interactions in neutral-atom quantum processors, and individual operations as digital steps to implement the protocol. To further improving the detection of complex features, we apply multiple quantum kernels by varying the qubit connectivity according to the hardware constraints. An architecture that combines non-trainable quantum kernels and standard convolutional neural networks is used to classify realistic medical images, from breast cancer and pneumonia diseases, with a significantly reduced number of parameters. Despite this fact, the model exhibits better performance than its classical counterparts and achieves comparable metrics according to public benchmarks. These findings demonstrate the relevance of digital-analog encoding, paving the way for surpassing classical models in image recognition approaching us to quantum-advantage regimes.

\end{abstract}

\maketitle

\section{Introduction}

Convolutional neural networks (CNNs) are well suited for image classification tasks \cite{cnn1} because they can extract relevant information via convolutions. In this way, a trainable kernel, much smaller than the original image, scans across and generates a series of reduced-dimensionality images.

Recently, implementations of quantum convolutional neural network (QCNN) have been developed~\cite{qcnn1}. This model applies convolution and pooling layers within a digital quantum circuit, starting from an initial state that encodes classical information. QCNN has shown improvements over existing models in recognizing quantum phases and was applied to quantum error correction. Additionally, other QCNN types have been proposed for image classification~\cite{deep_qcnn_qst, qcnn_qmi, multiclass_qcnn_frontier}. These include models whose convolutions are carried out using a non-trainable quantum kernel, consisting of an encoding layer and a highly entangled circuit~\cite{qcnn2}. There has been keen interest in applying QCNNs to medical image recognition~\cite{qcnn_med_ieee, qcnn_covid_1}. In these models, entanglement enhances the model's expressivity~\cite{expressivity, expressivity_nature}, which has been shown to saturate as the number of layers increases. However, the practical implementation of these models with digital quantum circuits, within the qubit coherence times, poses difficult challenges in the current era of noisy intermediate-scale quantum (NISQ) processors.

The digital-analog quantum computing (DAQC) paradigm was proposed as a means to harness the strengths of both digital and analog quantum computing. This is done by combining the flexibility of digital quantum gates with the robustness of multipartite-entangled analog blocks~\cite{daqc, sc_daqc, daqc_fourier_transf, daqc_superconducting}. These analog blocks are derived from the native interactions of the available quantum hardware, so that the coherence time is efficiently used. The DAQC approach involves decomposing the problem into components as digital gates, performing single-qubit rotations, and analog blocks executing multi-body quantum interactions through time-dependent couplings. This method has been implemented across various quantum processors, including superconducting circuits, trapped ions, and neutral-atom quantum simulators for diverse applications, demonstrating superiority over the traditional gate-based model~\cite{qml_daqc_sc, quera_qml, daqc_ions_sr, daqc_trapped_ions, daqc_ions_aps, variational_daqc_nn}.
In the last years, neutral-atom platforms have showcased a scalable approach to building programmable quantum processors consisting of hundreds of qubits \cite{na_daqc, na_daqc_2}. These devices have been used for studying combinatorial optimization problems~\cite{mis_science, mis-arxiv-2018}, quantum simulations~\cite{q_sim_na_nature}, learning tasks~\cite{quera_qml}, among others~\cite{na_daqc}. Furthermore, these devices are adequate for DAQC, where the time-dependent Hamiltonian of the analog block is given by
\begin{equation}
    \frac{\mathcal{H}(t)}{\hbar} = \frac{\Omega(t)}{2}\sum_{j}\sigma_j ^x - \delta(t)\sum_{j}\eta_j + \sum_{<i,j>}J_{i,j} \eta_i \eta_j
    \label{ryd_ham} .
\end{equation}
Here, $\sigma^{x,z}$ are Pauli matrices and $\eta_i = \left(1 -\sigma^z _i\right)/2$. The interaction strength is $J_{i,j}=C_6/r_{i,j}^6$, where $C_6$ is the constant depending on the device, and $r_{i,j}$ is the distance between the two atoms $i$ and $j$. By combining these available analog blocks in neutral-atom quantum computers with digital gates, one can engineer quantum algorithms for a given problem of interest.

This work proposes the hybrid composition of standard CNN  mehods with non-trainable digital-analog quantum kernels for image classification. In this setup, digital gates are used to represent $n \times n$ pixels from the $N \times N$ original ones. This is followed by a time-dependent analog block with $n^2$-qubits to perform convolutions through the image, while observables are measured to encode information for generating new images. The images generated from the convolutions should extract relevant, complex features and correlations between pixels, which are then used to feed a classical model with trainable parameters. Additionally, multiple topologies of digital-analog quantum kernels are proposed by varying the connectivity between qubits to capture a greater number of features between pixels. Simulations were conducted using two realistic datasets of medical images~\cite{medmnist} for breast cancer and pneumonia classification. The presented results show that quantum kernels using digital-analog encoding, even if non-trainable, can detect features in the image that are more beneficial for the models than the most advanced classical convolutional layers. The area under the Receiver Operating Characteristic (ROC) curve and the corresponding accuracy were used as benchmark metrics.

\section{Digital-analog quantum convolutional neural networks (DAQCNN)}

In order to perform convolutions through digital-analog quantum kernels (DAQKs), the classical information contained in each portion of the image, $K_{n\times n}$, should be efficiently encoded into the quantum circuit. For this, one applies angle embedding, in which a flattened feature vector, $\mathbf{\phi}$, containing information about the pixels, is encoded into single-qubit rotations. Fig.~\ref{quanvolution} shows a digital block that is applied on a zero-state initialization. For this work, we consider the single-qubit rotations as $H R_y(\phi_{i,j})$, encoding pixel information, $\phi_{i,j}$, on a quantum state, $H R_y(\phi_{i,j})|0\rangle = |\phi\rangle_{i,j}$, given by
\begin{equation}
    |\phi\rangle_{i,j} = 
    \frac{\left(\cos{\frac{\phi_{i,j}}{2}} + \sin{\frac{\phi_{i,j}}{2}}\right)|0\rangle + \left(\sin{\frac{\phi_{i,j}}{2}} - \cos{\frac{\phi_{i,j}}{2}}\right)|1\rangle}{\sqrt{2}}.
    \label{init_state}
\end{equation}
The Hadamard gate, $H$, is used to minimize the likelihood of the projection in the $z$-basis. This type of rotation is suitable when the angle embedding is followed by interaction blocks of the type $e^{-i\theta\sigma^z_i\sigma^z_j}$ and there exist angles such that $\phi_{i,j}=0$.
A highly entangled analog block, whose evolution is given by
\begin{equation}
    U(\tau) = \mathcal{T} e^{-i\int_{0}^{\tau}\mathcal{H}(t)dt},
    \label{eq_analog}
\end{equation}
is then applied in order to capture complex relationships among the encoded pixels. The entangled state produced is given by $U^\mathbf{g}(\tau)|\mathbf{\phi}\rangle = |f\left(\mathbf{\phi}, \tau, \mathbf{g}\right)\rangle$, where $\mathbf{g}$ is a graph containing a feasible connectivity according to the hardware of interest. Measurements of observables are performed to extract information which will be encoded into the generated pixel values. The next subsections are meant to explain two approaches designed for this work, whose core difference is that the first uses a single quantum kernel to perform convolutions and the latter uses multiple quantum kernels, by changing the connectivity of $\mathbf{g}$, to perform the same task. The final goal is to propose DAQCNN as a successful merge of DAQC methods with standard CNN protocols for image classification.

\subsection{Single and multiple digital-analog quantum kernels}

\begin{figure*}
    \centering
    \includegraphics[width=1 \linewidth]{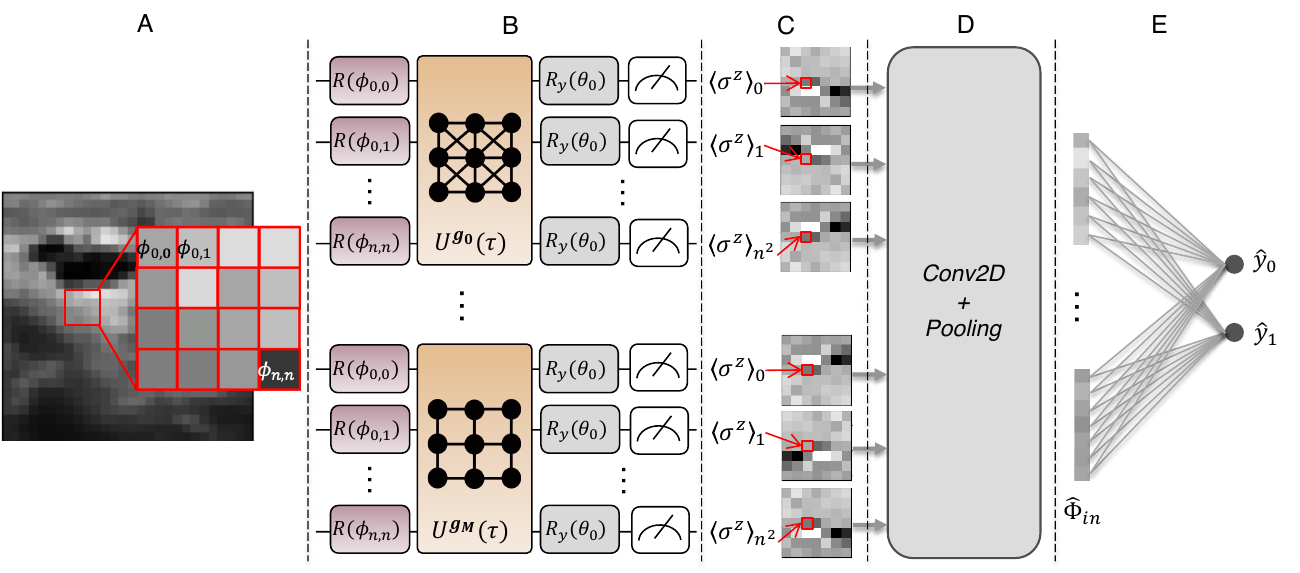}
    \caption{Convolutions with single-DAQKs and multi-DAQKs. The figure illustrates a DAQCNN applied to the original image. (A) the original image sourced from Ref.~\cite{medmnist} is displayed, with highlighted target pixels denoted as $\Phi = \{\phi_{0,0}, \phi_{0,1}, ..., \phi_{n,n}\}$, where the convolution will take place. (B) These $n^2$ pixels undergo DAQKs composed of single-qubit rotations, $R(\phi_{i,j})$, analog blocks, $U^{\mathbf{g_j}}(\tau)$, and measurements in the computational basis. One or more analog blocks $U(\tau)$ (see Eq. \ref{eq_analog}) with different graph connectivity, $\mathbf{g}$, can be applied, single- and multi-DAQK, respectively, followed by an additional digital block with a global $R_y(\theta_0)$ rotation. (C) Expectation values of single-body observables, $\langle \sigma^z \rangle_i$, are computed, which will be utilized in generating new images with reduced dimensions. (D) The quantum-generated images go through standard convolutional and pooling layers. (E) The resulting images are flattened as a input vector, $\hat{\Phi}_{in}$, and passed to a dense trainable layer whose outputs, $\hat{y}_0$ and $\hat{y}_1$, estimate the target binary variables, as an example of binary classification problems}
    \label{quanvolution}
\end{figure*}

In the case of single-DAQK, a unique quantum kernel is used to perform the convolutions on the original image. As seen in Fig.~\ref{quanvolution}, the $n^2$ pixels from the portion $K_{n\times n}$ are normalized and encoded into single-qubit rotations and one analog block for a particular graph, $\mathbf{g}$, is applied. Note that only one graph from Fig.~\ref{quanvolution} should be considered in this case. An expectation value from an observable, $\sigma^z_i$, is measured from each $i$-th qubit. The observations, $\langle \sigma^z_i \rangle$, are encoded into new images, which are responsible to detect complex features from the original image. Thus, the flattened output images for the single-DAQK approach, $\psi_{in}^s$, can be written as the linear combination
\begin{equation}
    \psi_{in}^s = \sum_{p = i,j \in K_{n\times n}} \langle\phi|_pU^\mathbf{g}(\tau)^\dagger \sigma^z U^\mathbf{g}(\tau)|\mathbf{\phi}\rangle_p \cdot \hat{e}_p .
    \label{out_single}
\end{equation}
Here, $\hat{e}_p$ represents the basis and the notation $\psi_{in}^s$ is chosen according to the Fig.~\ref{quanvolution}, whose output from the DAQK is the transition from C to D. 

In order to detect even more characteristics from the original images, we propose an approach where one implements multi-DAQK, by varying the topology of $\mathbf{g}$. We introduce $M$, which is the number of different graphs. In this case, the difference between the multiple kernels is given by their connectivity graphs, $G = \{\mathbf{g_0}, \mathbf{g_1}, ..., \mathbf{g_M}\}$, as also illustrated in Fig.~\ref{quanvolution}. In this case, similarly to Eq.~(\ref{out_single}), the flattened output vector can be written as 
\begin{equation}
    \psi_{in}^m = \sum_{m=1}^{M}\sum_{p = i,j \in K_{n\times n}} \langle\phi|_{p,M}U^\mathbf{g}(\tau)^\dagger \sigma^z U^\mathbf{g}(\tau)|\mathbf{\phi}\rangle_{p,M} \cdot \hat{e}_{p,M} .
    \label{out_multiple}
\end{equation}

In the single-DAQK approach, the maximum number of quantum-generated images is restricted by the number of qubits, $n^2$, while this number is $Mn^2$ for multi-DAQK. Both approaches provide augmented information from the data set of interest. Multi-DAQK can detect more characteristics from the original image, thus producing a more representative input for the classical models. Nonetheless, this approach requires more hardware evaluations, which can be a drawback depending on the particular runtime of the hardware where the problem is being tackled. This problem can be overcome through the utilization of on-chip parallel processing.

To tackle high-resolution image classification problems spanning from $516\times 516$ to $1024\times 1024$, larger kernel size ranging from $8\times 8$ to $10\times 10$ is required, demanding $64$ to $100$ qubits, respectively. Our estimation shows that to enable parallel computations of DAQCs, tens of thousands of qubits would be essential. 
Moreover, the execution time of DAQCs, from operations to measurement results, needs to be approximately substantially faster than the current commercial hardware, typically around the order of 500ms. The work proposed in Ref.~\cite{na_imaging_us} would allow for an execution time in the range of microseconds, making it feasible to use DAQK in large-scale data. Alternatively, exploring hardware platforms with significantly reduced execution times, such as neutral atoms with hyperfine qubits, superconducting circuits, and trapped ion systems, could be viable solutions.

\subsection{Capacity of digital-analog quantum kernels}

While CNNs are complex systems that can behave unpredictably with respect to certain parameters, we can estimate their ability to deal with complex data through some quantities such as capacity and expressivity. The capacity and learnability of a model can be assessed using the Vapnik–Chervonenkis dimension (VC dimension) \cite{capacity_0, capacity_1, vc_dim_nn}, in addition to other metrics \cite{capacity_nn_NN}. These metrics collectively evaluate the model's effectiveness in fitting various functions. Similar works have been done to demonstrate the exprevissity of quantum models~\cite{power_qnn, expressivity_1, expressivity}. This degree naturally amplifies as the model's complexity expands, achieved through elevating entanglement levels alongside parametrised qubit rotations. In neural networks, complexity can be expressed in terms of their depth or number of neurons, leading to an increase in the number of parameters. The concept of local effective dimension as a capacity measure for machine learning models, demonstrating strong correlation with generalization error on standard datasets, has been proposed \cite{quantum_capacity_abbas}. To provide an explanation of why models using DAQKs exhibit better generalization than their purely classical counterparts for complex data sets, we will demonstrate that quantum models have exponentially greater capacity than classical models. Therefore, they cannot be simulated as the (number of qubits) kernel size scales.

The measured expected values, $\langle \sigma^z _i \rangle$ - if the initialization used is given by Eq.~(\ref{init_state}) - in the absence of entanglement would be explicitly given by $\langle \sigma^z \rangle_i = 2\cos(\frac{x_i}{2})\sin(\frac{x_i}{2})$, which we denote as order $\mathcal{O} = 1$, since it consists only of periodic functions in its expansion. It is clear that the absence of entanglement causes $\langle \sigma^z  \rangle_i$ to depend only on the rotations performed on the $i$-th qubit. In a system where a high degree of entanglement is incorporated, such as in DAQK, $\langle \sigma^z \rangle_i$ becomes a function that depends on all $n^2$ rotations encoding the input pixels. Each $\langle \sigma^z \rangle_i$, or in other words, each new pixel calculated in the convolutions, contains information about all other pixels in an exponential combination of periodic functions given by
\begin{equation}
    \langle\phi|_iU^\mathbf{g}(\tau)^\dagger \sigma^z U^\mathbf{g}(\tau)|\mathbf{\phi}\rangle_i = \sum_{r=1}^{\mathcal{O}} \alpha_{i,r}\prod_{j=1}^{\lceil M n^2 \rceil}p(x_j),
\end{equation}
where $p(x_j)$ are periodic functions derived from single-qubit rotations with the normalized pixels, $x_j$, and $\alpha_{i,r}$ are the coefficients which can be written as a DAQK-parameter function, $\alpha_{i,r}\left(\mathbf{x}, \tau, \mathbf{g}\right)$. The level of entanglement in $\mathbf{g}$ causes the number of terms in $\langle \sigma^z \rangle_i$ to grow exponentially with the number of qubits, $n^2$, namely $\mathcal{O} = \exp{(n^2)}$, encoding a dense mixture of information about all pixels encoded in the entangled qubits. This information will be implicitly carried to the output layer of the classical model. Therefore, we can assert that models using DAQK for convolutions exhibit exponentially higher capacity compared to those utilizing standard convolutions, which involve elementwise multiplication followed by non-linear activation functions. When increased capacity in models is important for their generalization ability, the use of DAQK emerges as a candidate to surpass classical models. 

\section{Computational experiments}

\begin{figure}
    \centering
    \includegraphics[width=0.9 \linewidth]{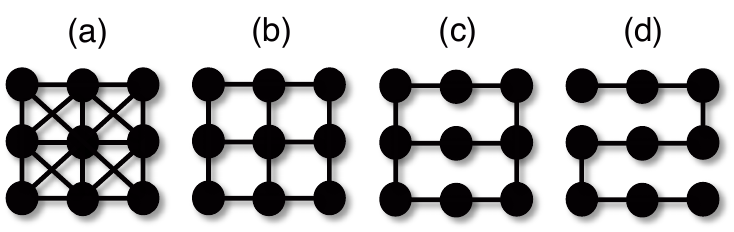}
    \caption{Graph topologies, $\mathbf{g}$, for analog blocks. Graph (a) represents the King's graph an the others, (b), (c), and (d), are variations derived from (a), according to neutral-atom hardware limitations.}
    \label{graphs}
\end{figure}

\begin{figure}
    \centering
    \includegraphics[width=\linewidth]{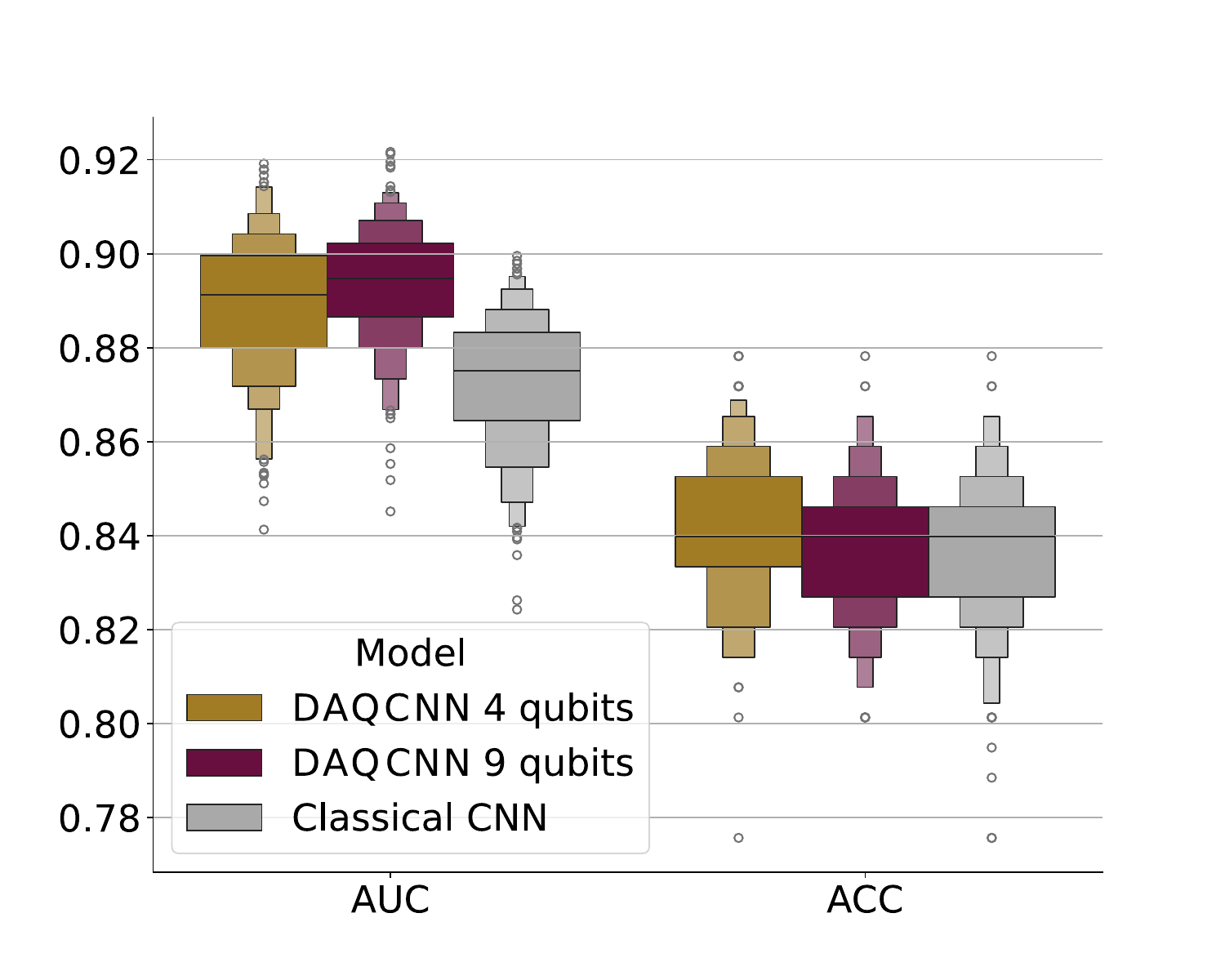}
    (a) Results for BreastMNIST
    \includegraphics[width=\linewidth]{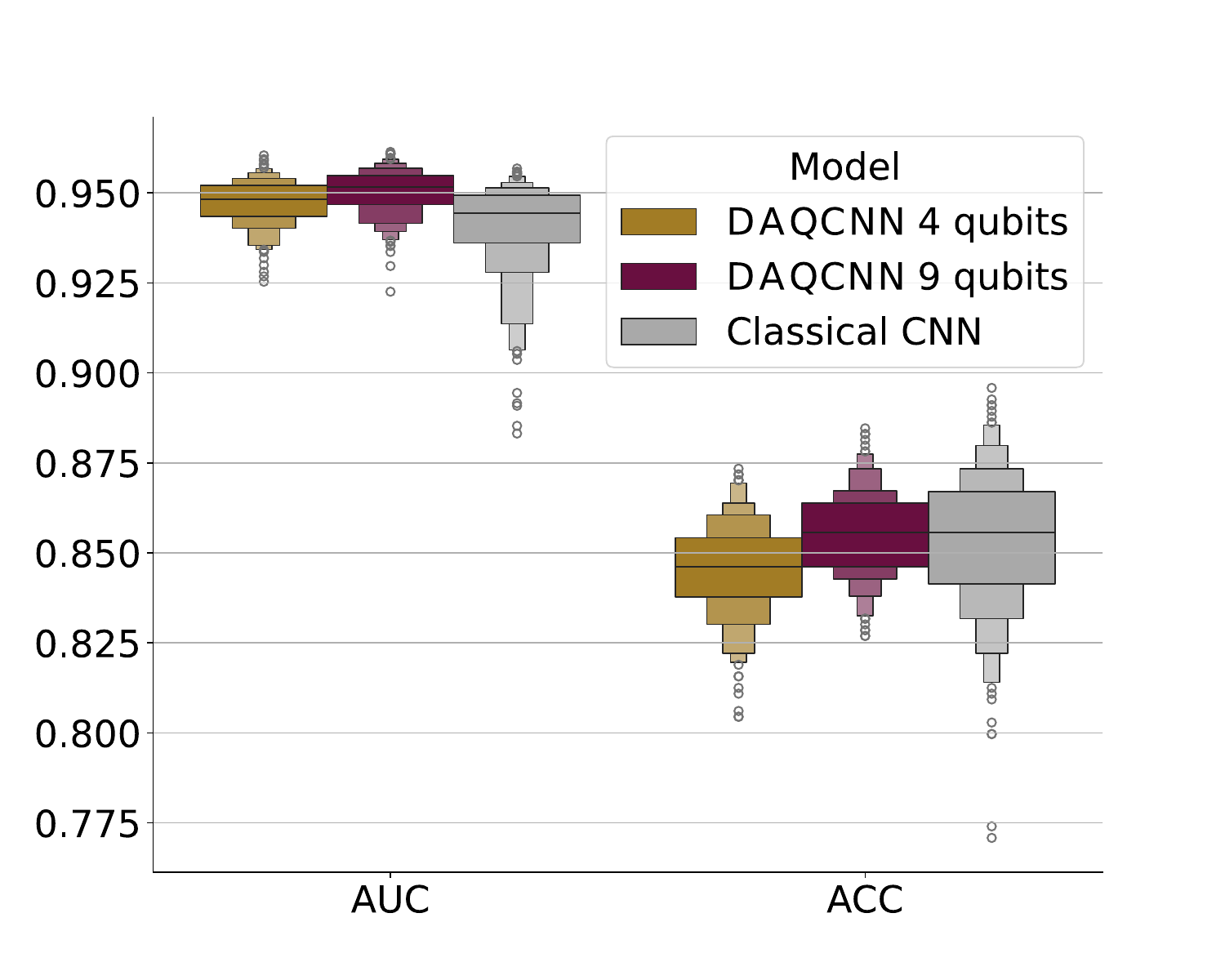}
    (b) Results for PneumoniaMNIST
    \caption{Statistical results for AUC and test ACC by varying the hyperparameters as well as the weight initialization. The models evaluated are DAQCNN 4 qubits, DAQCNN 9 qubits ($2 \times 2$ and $3 \times 3$ kernels, respectively) and the classical equivalent model, which has the same architecture and number of parameters. (a) shows the results for BreastMNIST, while (b) corresponds to the results on PenumoniaMNIST}
    \label{boxen_breast}
\end{figure}

To conduct computational experiments, we considered single- and multi-DAQK (with 1 and 4 kernels, respectively) with kernel sizes of $2 \times 2$ (4 qubits) and $3 \times 3$ (9 qubits). The four graph topologies used on the simulations are explicitly illustrated in Fig.~\ref{graphs}, where (a) represents the King's graph. We tested two datasets of medical images for binary classification, available at \cite{medmnist}, for breast cancer and pneumonia, respectively. We used a standard CNN model as a benchmark, with equivalent architecture and number of parameters (see Fig.~\ref{boxen_breast}). We also compared the performance of our models with large models from a publicly available benchmark \cite{medmnist}.

For simulating digital-analog quantum kernels (DAQK), we used the following parameters: the single-qubit rotations represented by $R(\phi_{i,j})$ in Fig.~\ref{quanvolution} were performed as $HR_y(\phi_{i,j})$, where $H$ is a Hadamard gate and $\phi_{i,j}$ is the normalized value of the pixel at position $(i,j)$, normalized such that $\phi_{i,j} = [0, \pi]$. The analog block, in turn, was simulated using trotterization of Eq.~\ref{eq_analog}, using the Rydberg Hamiltonian from Eq.~\ref{ryd_ham}. The time-dependent functions contained in Eq.~\ref{ryd_ham} were defined as $\Omega(t) = \delta(t) = t$. The total evolution time was set as $\tau = 0.2$, with trotter steps performed every $\delta t = 0.05$ (4 trotter steps). For cases where the single-DAQK approach was used, the qubit connectivity, $\mathbf{g}$, utilized was the King's graph, whose topology has the maximum connectivity in the current generation of neutral atom quantum hardware. In the multi-DAQK approach, variations of $\mathbf{g}$ respecting the native hardware interactions were used. A description of the architecture of the standard CNNs that process the quantum-generated images is shown in Table~\ref{tab:model_structure}.

\begin{table}[h]
\caption{CNN model architecture composed of multiple convolutional layers, batch normalization layers, max-pooling layers, dropout layers, and a densely connected layer. Each layer is described with its type and corresponding output shape, along with the number of trainable parameters. Furthermore, hyperparameters such as learning rate, activation function, and dropout rate have been carefully tuned to optimize the model's performance for each specific task.}
\centering
\begin{tabular}{ccc}
\hline
\hline
\textbf{Layer}           & \textbf{Output shape}  & \textbf{No. of parameters} \\ \hline
Conv2D           & ( 27, 27, 64)     & 320               \\ 
BatchNormalization & (27, 27, 64) & 256  \\ 

MaxPooling2D & (13, 13, 64)    & 0                 \\ 
Conv2D            & (12, 12, 64)     & 16448             \\ 
Dropout         & (12, 12, 64)     & 0                 \\ 
Flatten           & (9216)           & 0                 \\ 
Dropout          & (9216)           & 0                 \\ 
Dense               & (1)              & 9217              \\ 
\hline
\hline
\end{tabular}
\label{tab:model_structure}
\end{table}

Statistical results showcasing the impact of varying hyperparameters and parameter intialization are provided in Fig.~\ref{boxen_breast}. The datasets used in this analysis are tailored for classifying cases of breast cancer and pneumonia. The three models share the same architecture with a closely similar number of classical trainable parameters ($\sim30\times 10^3$). A grid search, coupled with parameter initialization, was conducted on all three models to determine the optimal hyperparameters and parameter initialization settings. The grid search encompassed hyperparameters including learning rates \{0.0001, 0.0005\}, dropouts \{0.55, 0.6\}, and activation functions \{ReLU, GELU\}. We initialized the model $30$ times in order to demonstrate the robustness of the DAQCNN and its superiority over its classical counterpart. While the models may demonstrate comparable overall test accuracy (ACC), it is crucial to underscore that the datasets exhibit class imbalance (approximately $75/25\%$). Therefore, AUC emerges as a more appropriate metric for robust evaluation.

\begin{table}[htbp]
    \centering
    \caption{Area Under ROC Curve (AUC) and accuracy (ACC) for test and validation set, evaluated over breast cancer and pneumonia classification problems~\cite{medmnist}. Best results are highlighted in bold.}
    \begin{tabular}{cccccc}
        \hline
        \hline
        \multirow{2}{*}{Dataset} & \multirow{2}{*}{Model} & \multicolumn{2}{c}{Test} & \multicolumn{2}{c}{Validation} \\
        \cline{3-6}
         & & AUC & ACC & AUC & ACC \\
        \hline
        \multirow{5}{*}{Breast} & DAQCNN 2x2 (4 graphs) & 0.918 & 0.871 & 0.910& 0.910 \\
         & DAQCNN 2x2 (1 graph) & 0.919 & 0.878 & 0.884 & 0.885 \\
         & DAQCNN 3x3 (4 graphs) & \textbf{0.926} & \textbf{0.891} & \textbf{0.923} & \textbf{0.936}\\
         & DAQCNN 3x3 (1 graph) &  0.919 & 0.858 & 0.894 & 0.898\\
         & Equivalent CNN & 0.903 &0.885 &0.913 & 0.923\\
         & ResNet 18 & 0.901 & 0.863 & - & - \\
         & ResNet 50 & 0.866 & 0.842 & - & - \\
         & Google Vision & 0.919 & 0.861 & - & -\\
        \hline
        \multirow{5}{*}{Pneumonia} & DAQCNN 2x2 (4 graphs) & 0.960 & 0.861 & 0.996 & 0.969 \\
         & DAQCNN 2x2 (1 graph) & 0.962 & 0.864 & 0.996 & 0.966 \\
         & DAQCNN 3x3 (4 graphs) & 0.962 & 0.884 & 0.996 & 0.968\\
         & DAQCNN 3x3 (1 graph) & 0.963 & 0.858 & \textbf{0.997} & 0.964 \\
         & Equivalent CNN & 0.957 & 0.849 & \textbf{0.997} & \textbf{0.975} \\
         & ResNet 18 & 0.956 & 0.864 & -&- \\
         & ResNet 50 & 0.962 & 0.884 & - & - \\
         & Google Vision & \textbf{0.991} & \textbf{0.946} & -&- \\
        \hline
        \hline
    \end{tabular}
    \label{tab:experiment_results}
\end{table}

Table~\ref{tab:experiment_results} presents a comparison of the best results obtained for DAQCNN models with different kernel sizes and state-of-the-art classical models. Public results from large models, found in \cite{medmnist}, have also been included. It is relevant to note that ResNet 18, ResNet 50 and Google Vision reference models contain a significantly larger number of parameters ($\geq 11\times 10^6$) compared to our DAQCNN and its equivalent standard CNN. We addressed two binary classification tasks, focused on breast cancer and pneumonia. It is remarkable that the classical models used in \cite{medmnist} exhibit satisfactory performance for PneumoniaMNIST. Consequently, there may be limited scope for improvement in this scenario, and the introduction of additional complexity, such as with DAQCNN, may not yield discernible benefits. In any case, it is evident that for both datasets, models employing DAQCNN outperform their classical CNN counterpart.
The results for ResNet 18 and Google Vision were extracted from the public benchmark~\cite{medmnist}. ResNet 18, ResNet 50 and Google Vision are large models whose number of parameters are $\geq 11\times 10^6$. The four DAQCNN models presented, along with their classical counterpart referred to as Equivalent CNN, use up to $30\times 10^3$ parameters. To achieve these results, we conducted a grid search for hyperparameter optimization, running each model $30$ times and selecting the best-performing values.

\section{Discussion and conclusion}

We have proposed DAQCNN as a powerful paradigm for image classification. The proposed approach combines standard CNNs with non-trainable DAQKs, leveraging the benefits of both digital and analog quantum computing paradigms. By encoding image information into quantum states and exploiting complex entanglement dynamics, DAQKs can capture intricate features and correlations within realistic images.

The DAQCNN results demonstrate promising performance of the DAQK-enhanced models compared to their classical counterparts. Both single- and multi-DAQK approaches show improvements in classification metrics such as Area Under ROC Curve (AUC) and test accuracy (ACC) when applied to datasets of breast cancer and pneumonia classification. Remarkably, the DAQK-augmented models achieve comparable results to large-scale classical models, showcasing the potential of quantum-enhanced methodologies in image recognition tasks. Through the incorporation of entanglement dynamics, DAQKs offer exponentially greater capacity compared to classical convolutional layers, allowing for the extraction of complex features and correlations from images. 

The choice of qubit connectivity graphs and optimization of quantum circuit parameters play a crucial role in harnessing the full potential of DAQKs. The proposed multi-DAQK approach, which explores variations in connectivity graphs, offers a more comprehensive exploration of image features, although it leads to an increased computational overhead.

Further research will delve into the scalability and applicability of DAQCNN to larger and more diverse datasets, along with implementations on various types of quantum hardware, such as trapped ions and superconducting circuits. Significant enhancements can be anticipated upon the next generations of analog blocks overcoming hardware limitations, such as leveraging ground-Rydberg qubits with time-dependent local control, manipulation of interactions, or employing hyperfine qubits. Additionally, advancements in quantum computing platforms will be pivotal in unlocking the full potential of DAQCNN in real-world image recognition tasks. In neutral-atom quantum processors, the primary challenge lies in overcoming the execution time of a DAQC, presently confined to the order of 500 ms. This challenge could be tackled by substantially increasing the number of qubits, thus enabling efficient parallelization of DAQC operations.

\begin{acknowledgments}
This work has been partially supported by the Valencian Government grant  CIAICO/2021/184 and the Spanish Ministry of Economic Affairs and Digital Transformation through the QUANTUM ENIA project call – Quantum Spain project, and the European Union through the Recovery, Transformation and Resilience Plan – NextGenerationEU (Digital Spain 2026) Agenda. 

\end{acknowledgments}

\bibliography{reference.bib}

\end{document}